\documentstyle[aps,preprint]{revtex}
\newcommand{\be}{\begin{equation}}
\newcommand{\ee}{\end{equation}}
\newcommand{\ba}{\begin{eqnarray}}
\newcommand{\ea}{\end{eqnarray}}
\newcommand{\RN}{Reissner-Nordstr\"om}
\def\tr{\mathop{\rm tr}\nolimits}
\date{\ }
\begin{document}
\preprint{To appear in Phys. Rev. D}
\title{Spontaneous loss of charge of the \RN\, black hole}

\author {Cl. Gabriel
 \thanks{ Aspirant F.N.R.S.}}

\address {Dept. of Mechanics and Gravitation, University
of Mons-Hainaut, Mons, Belgium}

\date{\today}

\maketitle
\begin{abstract}
In this paper, we study by a functional method the vacuum instability of a charged scalar field,  when it is quantized in the background of the \RN\, black hole; we also show that the first stage of the evaporation process of the black hole can be driven by a Schwinger-like effect.
\end{abstract}

\section{Introduction}
Since Hawking's seminal result \cite{Hawking}, it's well known that black hole evaporate by emitting a thermal flux of particles. However, for a black hole carrying an electric charge, it's natural to expect that the hole will preferentially emit particles of the same charge as its own charge, because of the electrostatic repulsion. So the hole must spontaneously lose its charge. This result was first established by Gibbons in  \cite{Gibbons}. In his paper, Gibbons used classical methods of quantum field theory in presence of external fields, as the building of $in$ and $out$ states, and the computation of the Bogoljubov transformation connecting the two bases of states. In our paper we obtain the same conclusion as Gibbons by using a functional method, which doesn't need the explicit evaluation of the bases and the Bogoljubov transformation, which is very hard in the black hole case.\\
A non rotating charged black hole, of mass $M$ and charge $Q$ is described in $3+1$ dimensional spacetime by the \RN\, metric:
\ba
ds^2=- \Delta \, dt^2-\Delta^{-1} dr^2+r^2(d\theta^2+\sin^2\theta d\varphi^2)\label{RN}\quad,
\ea
where $\Delta$ is the function:
\ba
\Delta=1-\frac{2M}r+\frac{Q^2}{r^2}\qquad .
\ea
In this paper, I study only non extremal charged black hole, that's to say black hole such that $M^2>Q^2$.\\
This metric possesses two horizons : an event horizon for $r=r_+=M+\sqrt{M^2-Q^2}$ and an inner horizon in $r=r_-=M-\sqrt{M^2-Q^2}$.
The electric field of the hole can be derived from the vector potential:
\ba
A_\mu dx^\mu=-\frac Q r dt\label{RNelec}.
\ea
Following \cite{Gibbons}, we expect that some of these charged black hole will spontaneously lose their charge by radiating charged particles of the same sign as their own charge $Q$. This phenomenon happens specially when the hole is such that $Mm \ll Qq$ \footnote{where $m$ and $q$ are the mass and the charge of the lightest charged particle pair (the electron-positron pair)}, that's to say if the electromagnetic Coulomb interaction exceeds the Newtonian  gravitational attraction. As $r_+ > M$, this inequality can be written $qQ/r_+ \gg m$. We only consider below black holes that satisfy this inequality. Under this last condition, when a pair particle-antiparticle spontaneously appears in the vicinity of the hole, it's energetically favourable for a pair of oppositely charged particles to form in the vicinity of the black hole, that the black hole ``absorbs" the antiparticle and ``emits" the particle \cite{Gibbons}. In quantum field theory, this process causes vacuum instability for a charged field, quantized around the black hole.
\section{Functional integral and transition ratio}
Let us quantize a charged scalar field $\phi$ on the \RN\, background spacetime. Vacuum instability occurs from the existence of gravitational and electric fields of the black hole, and we must introduce $in$ and $out$ states for the quantized field. The vacuum persistence amplitude $\langle 0,out\vert 0,in\rangle$ can be obtained from  ``mode"
calculations, as in Gibbons's work, but may also be derived from the Green functions of the
quantized field (see for instance the book
\cite{BD}). Indeed, as  it is well known, the vacuum persistence amplitude can be obtained from the functional ${\cal{Z}}\left[J,J^*\right]$ as:
 \begin{eqnarray}  \langle 0,out\vert
0,in\rangle_{J=J^*=0}={\cal{Z}}\left[0,0\right]
\end{eqnarray} where ${\cal{Z}}[J,J^*]$ is defined by the functional integral~:
\begin{eqnarray}
{\cal{Z}}\left[J,J^*\right]&=&\equiv e^{i{\cal{W}}}=\int{\cal{D}}\phi{\cal{D}}\phi^*\,
\exp\left\{iS\left[\phi,\phi^*\right]+i\int J^*\phi d^4 x+i\int J \phi^*
d^dx \right\}\nonumber\\
\end{eqnarray} 
expressed in terms of the action $S\left[\phi,\phi^*\right]$ of the
charged scalar field, minimally coupled to the electric field:
\begin{eqnarray}  S\left[\phi,\phi^*\right]&=&-\frac12\int d^4x
\left\{({\cal{D}}_\mu\phi) ({\cal{D}}^\mu\phi)^*+({\cal{D}}_\mu\phi)^*
({\cal{D}}^\mu\phi)+\right.\nonumber\\
&&\left.(m^2-i\epsilon)\phi
\phi^*+(m^2+i\epsilon)\phi^*\phi)\right\}\nonumber\\ &=&-\int dx
dx'\phi^*(x) {\cal{H}}_{xx'}\phi(x')\qquad.
\end{eqnarray}
\\
A standard computation \cite{BD} allows us to give a meaning to the expression of $\cal{W}$~: 
\begin{eqnarray}  {\cal{W}}=-i{\tr}\ln(-{\cal{G}}) =\int_{m^2}^\infty
dm^2 \int d^4x \, G_F(x,x)
\end{eqnarray}  where ${\cal{G}}=-{\cal{H}}^{-1}$ is related to the
Feynman propagator
\begin{equation} G_F(x,y)=-i {\langle 0,out\vert{\cal{T}}
(\phi(x)\phi^\dagger(y))\vert 0,in
\rangle}/{\langle 0,out\vert 0,in\rangle}
\end{equation}
 by:
\begin{eqnarray} {\cal{G}}(x,y)=\frac12 \left\{
G_F(x,y)+G_F^*(y,x)\right\}\qquad.
\end{eqnarray}
Thus, the imaginary part (the one which encodes the vacuum instability)
of
 $\cal W$ reads:
\begin{eqnarray}  {\cal{I}}m{\cal{W}}={\cal{R}}e\int_{m^2}^\infty dm^2
\int d^4x \, {\cal{G}}(x,x)={\cal{R}}e\int_{m^2}^\infty dm^2 \int d^4x \,
G_F(x,x)\qquad.
\end{eqnarray} 
We now briefly recall the so-called Schwinger
representation of the Feynman propagator. Formally,  the Feynman
propagator $G_F$ is the inverse of the kinetic operator $K=\frac
{\delta^4(x-y)}{\sqrt{-g(x)}}\left\{-{\cal D}_\mu {\cal D}^\mu
+(m^2-i\epsilon)\right\}$ and can thus be computed as $G_F=-K^{-1}=-i \int_0^{\infty}ds e^{-i K s }$.
To express it, Schwinger \cite{Schw} introduced  the
kernel  defined as
$K(x,y;s)=\langle x^\prime\vert e^{-i K s} \vert x\rangle
$ which  obeys a Schr\"odinger  equation~:
\begin{eqnarray}
({\cal D}_\mu {\cal D}^\mu
-m^2+i\frac{\partial}{\partial s})K(x,y;s)=0
\end{eqnarray} with the boundary condition $K(x,y;s\rightarrow
0^+)=\delta^4(x-y)$.
 Using this ``proper time" representation of the Feynman propagator
 \begin{equation}
 G_F(x,x')=\int_0^\infty ds\,K(x,x';s)\qquad ,\label{Schwprop}
 \end{equation} the vacuum persistence amplitude can be written as:
\begin{equation}  {\cal{I}}m{\cal{W}}={\cal{I}}m\int_{m^2}^\infty dm^2
\int d^4x
\int_0^\infty ds\, K(x,x;s) \qquad .\label{WKint}
\end{equation} The kernel $K(x,x^\prime;s)$, defined as the matrix elements
of the evolution operator
$K$, can be expressed as a path integral:
\begin{equation}  K(x,x^\prime;s)=\int
{\cal{D}}X^{\mu}(s^\prime)e^{iS(x,x^\prime;X^\mu(s^\prime);s)}\label{Kkernel}
\end{equation} where the domain of integration covers all the paths $X^\mu(s^\prime)$ that
connect the point $x$ to the point
$x^\prime$ in a (rescaled) time\footnote{This time variable is
proportional to the proper time measured along the natural trajectory
that connects $x$ to $x'$, but, in general, does not coincide with it.}\  $s$,
and $S(x,x^\prime;X^\mu(s^\prime);s)$ is the classical action computed along these
trajectories. This action is given by: 
\begin{eqnarray}
S(x,x^\prime;X^\mu(s^\prime);s)=\int_0^s ds^\prime\left\{\frac{\dot X^2}4+q\dot
X^\mu A_\mu(X[s^\prime]) -m^2\right\}\qquad.
\end{eqnarray}
\section{Vacuum instability for the charged black hole}
In the charged black hole case, the classical action of a charged particle moving close to the \RN\, black hole is given by:
\ba
S=\int_0^sds^\prime&\left\{-\frac{\dot t^2}4\Delta+\frac{\dot r^2}4\Delta^{-1}+\frac {r^2}4\sin^2\theta \dot \varphi^2+\frac {r^2}4\dot \theta^2+\frac {r^2}4\sin^2\theta \dot \varphi^2+\right.\nonumber\\
&\left.\frac{qQ} r \dot t-m^2\right\}.
\ea
In the following, we only deal with a 2-dimensional reduced action, by omitting angular variables in the functional integration. The price to pay for this simplification is that the final pair production rate $\vert\langle 0,out\vert 0,in\rangle\vert^2$ will be proportional to the electric field of the hole (as for the case of Schwinger effect in 1+1 dimensional spacetime), rather than the square of the electric field (as in the case of the usual Schwinger effect in 1+3 dimensional spacetime) \cite{report}.\\
By techniques analogous to those worked out in the paper \cite{Rindler}, we can reduce the evaluation of the propagator (\ref{Kkernel}) to the computation of a one dimensional functional integral:
\ba
K(x,x^\prime;s)&=&\frac 1 r\int\frac{d\omega}{2\pi} e^{-i\omega(t-t^\prime)}
K^{RN}_\omega(r,r^\prime;s)\label{PropRN}\quad,
\ea
expressed in terms of the kernel:
\ba
K^{RN}_\omega(r,r^\prime;s)=\int {\cal{D}}r \exp \left\{i\int_0^s{\cal{L}}_\omega(r,\dot
r)ds^\prime\right\},\label{PropRNbis}
\ea
where the domain of the path integral is the set of all paths that connect the point $r$ to the point $r^\prime$ in a time $s$, and where the Lagrangian is given by: 
\ba
{\cal{L}}_\omega(r,\dot r)&=&\Delta^{-1}\left(\frac {\dot r^2}4+(\omega+\frac{qQ}r)^2\right)-m^2.\label{LagRN}
\ea
To estimate the vacuum persistence amplitude, we first use a quadratic approximation for the Lagrangian (\ref{LagRN}), in order to make trivial the path integral (\ref{PropRNbis}). We define the new radial coordinate $\rho$ by:
\ba
\rho(r)&=&\int_{r^+}^r \frac{dr^\prime}{\sqrt{\Delta(r^\prime)}}\nonumber\\
&=&r\Delta^{1/2}+M\mbox{\rm{Argcosh}}\,\frac{r-M}{\sqrt{M^2-Q^2}}\quad.
\ea
Then, the Lagrangian reduces to the expression:
\ba
{\cal{L}}_\omega&=&\frac 14\left(\frac{d\rho}{ds}\right)^2-V[\rho(r)]\quad,
\ea
where the potential is defined implicitly by:
\ba
V[\rho(r)]=-\frac 1\Delta (\omega+\frac{qQ}r)^2+m^2\label{potentielRN}\quad.
\ea
Let us compute the extrema of this potential; we get:
\ba
\frac{dV}{d\rho}&=&-\frac{2}{r^3 \Delta^{3/2}}\left[(\omega Q^2+MqQ)-r(M\omega+qQ)\right](\omega+\frac{qQ}r)\quad.
\ea
As $r_+>M$, the zero of the term in brackets is a maximum of the potential, but it belongs to the interval $[r_-,r_+]$ and is therefore always located under the event horizon $r=r_+$, that's to say outside of the domain of definition of the coordinates $(t,r)$. In fact, the potential has a maximum in the external region of the black hole ($r>r_+$) only for the frequencies $\omega$ in the interval $[-\frac{qQ}{r_+},0]$; this maximum is then located in $r_{max}=-qQ/\omega$. Only the frequencies that belong to this interval will contribute to the pair production amplitude. This behaviour is analogous to the vacuum instability of a charged scalar field, quantized from the point of view of a Rindler observer. Indeed, in this last case, not all the frequencies contribute to the vacuum instability \cite{Rindler}. A simple computation gives for the width of the potential at the maximum $r=r_{max}$:
\ba
\left.\frac{d^2V}{d\rho^2}\right|_{r=r_{max}}&=&-\frac{2\omega^4}{q^2Q^2}\quad.
\ea
Therefore, near the maximum $r=r_{\max}$, the potential can be approximated by the parabolic profile:
\ba
V(\rho)\simeq m^2-\frac{\omega^4}{q^2Q^2}(\rho-\rho_{max})^2\label{quadRN}.
\ea
By a straightforward calculation \cite{Rindler}, \cite{these}, we get for the imaginary part of the functional ${\cal W}$ the result:
\ba
{\cal{I}}m {\cal{W}}=\int dr I\label{trace}
\ea
with $I$ defined by:
\ba
I&=&\int_{-{qQ}/{r_+}}^0 d\omega \ln(1+\exp^{-\pi \frac{m^2 qQ}{\omega^2}}).\label{IntI}
\ea
To perform this last integration, which is a sum over the frequencies $\omega$ which contribute to the vacuum instability, we first note that the main contribution to this integral comes from the region $\omega\simeq\omega_{min}=-\frac{qQ}{r_+}$ and therefore, the result should be:
\ba
I&\simeq&e^{-\pi\frac{m^2r_+^2}{qQ}}\int_{-{qQ}/{r_+}}^0 d\omega=\left(\frac{qQ}{r_+}\right)e^{-\frac{\pi m^2r_+^2}{qQ}}\quad.
\ea
More precisely, making the change of variable $\omega=-\left(\frac{\pi m^2 qQ}{x}\right)^{\frac 12}$, the integral (\ref{IntI}) becomes:
\ba
I&=&\frac 12 (\pi m^2 qQ)^{\frac 12}\int_{{\pi m^2 r_+^2}/{qQ}}^\infty dx x^{-\frac 32}\ln (1+e^{-x})\nonumber\\
&=&\frac 12 (\pi m^2 qQ)^{\frac 12}\sum_{k=1}^\infty \frac{(-1)^{k+1}}{k}\int_{{\pi m^2 r_+^2}/{qQ}}^\infty dx x^{-\frac 32} e^{-kx}.
\ea
The main contribution comes from the $k=1$ term. Neglecting all the other terms of the serie, we get:
\ba
I&\simeq&\frac 12 (\pi m^2 qQ)^{\frac 12}\Gamma(-\frac 12,\frac{\pi m^2 r_+^2}{qQ})
\ea
in terms of the incomplete Gamma  function. Moreover, we choose to study black holes such that $qQ/r_+ \gg m$. Therefore, if we limit ourselves to large mass black holes, more precisely to black holes larger in radius than the Compton wavelength of the electron ($r_+>M\gg\hbar/m$) \cite{Gibbons}, the second argument of the Gamma function ${\pi m^2 r_+^2}/{qQ}$ is close to zero, and we can limit ourselves to the first term in the series development of the Gamma function:
\ba
\Gamma(\alpha,x)&=&\Gamma(\alpha)-\alpha^{-1}x^\alpha M[1,1+\alpha; x]\nonumber\\
&\simeq&\Gamma(\alpha)-\alpha^{-1}x^\alpha. 
\ea
We obtain in this way for the integral $I$ the result:
\ba
I&=&-\pi (m^2 qQ)^{\frac 12} +\left(\frac{qQ}{r_+}\right)e^{-\frac{\pi m^2 r_+^2}{qQ}}\nonumber\\
&\simeq&\left(\frac{qQ}{r_+}\right)e^{-\frac{\pi m^2 r_+^2}{qQ}}
\ea
as it was suspected above.\\
In order to obtain the probability of vacuum persistence, we must now take the trace (\ref{trace}). The integration on the time like coordinate gives as usual \cite{Rindler} a factor $T$, corresponding to the lapse of time during which the electric field of the hole remains appreciable. Finally, multiplying and dividing by a factor $r_+$, in order to make the electric field of the black hole to appear, we get for the probability of vacuum persistence the result:
\ba
\vert\langle 0,out \vert 0,in\rangle\vert^2 &=&e^{-2{\cal{I}}m({\cal{W}})}=\exp\left[-\alpha qEV e^{-\pi\frac{m^2r_+^2}{eQ}}\right]\label{InstvideRN}\quad,
\ea 
where $\alpha$ is some numerical factor and $V$ is the "2-volume" of the hole.\\ This result is the same as the vacuum instability that results from Schwinger effect in flat $1+1$ dimensional space \cite{Rindler}, if we choose for the constant value of the electric field the value of the electric field of the hole on the horizon $E=Q/r_+^2$. This shows that the loss of charge process of the hole is described by Schwinger effect, as in ordinary flat-space quantum eletrodynamics. This result can be understood by noting that in the large radius hypothesis adopted above, the event horizon of the hole is much larger than the Compton wavelength of the  emitted particles, and the radiated pairs can be viewed as being produced by the electic field of the hole, and not by the gravitational field \cite{HiscockWeems}. Our result is compatible with the usual rate of electron-positron pair creation; indeed, remember that we obtained it in a reduced 1+1 dimensional spacetime. A more completed calculation, involving the evaluation of a four dimensional path integral would give a factor $E^2$ rather than a factor $E$ and a four dimensional volume, as expected.\\
In order to describe the evaporation scenarii for the charged black holes, a more detailed analysis is needed \cite{HiscockWeems}, because the holes also radiate neutral particles with a thermal rate. The key parameters to describe this evolution are the charge/mass ratio ($Q/M$) and the mass ($M$) of the hole. In fact, for a given $Q/M$ ratio , the not too heavy black holes  experiment in a first stage an evaporation process driven by Schwinger effect, rather than by Hawking effect. They spontaneously radiate their own charge and, once they are neutralized, continue to evaporate according to the usual Hawking process. On the other hand, for the same $Q/M$ ratio charge, the holes more massive lose mass by the Hawking process, while they conserve the same electric charge (the most massive even become "extreme"), before that the neutralization process really happens \cite{HiscockWeems}.
\section*{Acknowledgments}
I thank Ph.Spindel for stimulating discussions and the Fonds National de la
Recherche  Scientifique (F.N.R.S.) for financial support.

\end{document}